\documentstyle[12pt]{article}

\begin{document}
\baselineskip=1.0cm
\title{Effective Chiral Theory of Mesons}
\author{Bing An Li
\\Department of Physics and Astronomy \\ University of Kentucky \\
Lexington, KY 40506, USA}
\maketitle
{\large\bf Introduction}\\
Before the advent
of $QCD$ the meson physics has been studied extensively since 60's.
\begin{enumerate}
\item Chiral symmetry, Current algebra, PCAC.
\item Vector Meson Dominance(VMD).
\item Goldstone theorem.
\item Weinberg sum rules.
\item KSFR sum rule.
\item $\pi\pi$ scattering.
\item $\pi^{0}\rightarrow\gamma\gamma$ and triangle anomaly.
\item Wess-Zumino-Witten Lagrangain.
\item Chiral perturbation theory.
\item Quark model leads to a energy scale: constituent quark mass.
\end{enumerate}
How can we have a theory to unify all these studies?
This theory should be $QCD$ inspired, self-consistent, and
phenomenologically successful. We have proposed a theory (B.A. Li,
Phys. Rev. {\bf D 52}5165-5183, 5184-5193(1995)).

{\large\bf Simulations of the meson fields}\\
The key point is how to bosonize $QCD$.
As a matter of fact, in the bosonization of 1+1 filed theory
\[{1\over \sqrt{\pi}}\partial_{\mu}\phi=\bar{\psi}\gamma_{5}\gamma_{\mu}
\psi.\]
Following this idea,
we propose to simulate the
meson fields by quark operators. For example,
\[\rho^{i}_{\mu}=-\frac{1}{g_{\rho}m^{2}_{\rho}}
\bar{\psi}\tau_{i}\gamma_
{\mu}\psi.\]
{\Large\bf Model independent test}\\
Using PCAC, current algebra, and this expression, in the limit          t
of $p_{\pi}\rightarrow 0$, we obtain
\[{1\over 2}f_{\rho\pi\pi}g_{\rho}=1,\]
\[A={2\over f_{\pi}}(1-{1\over 2\pi^{2}g^{2}})^{-{1\over 2}}(m^{2}_{a}
-m^{2}_{\rho}).\]

The octet
pseudoscalars are Goldstone bosons and we treat pseudoscalar mesons
differently from others.\\
{\Large\bf
realization of quark operator expressions of meson fields}\\

Using $U(2)_{L}\times U(2)_{R}$ chiral symmetry and the minimum
coupling principle,
the Lagrangian is constructed as
\begin{eqnarray}
{\cal L}=\bar{\psi}(x)(i\gamma\cdot\partial+\gamma\cdot v
+\gamma\cdot a\gamma_{5}
-mu(x))\psi(x)-\bar{\psi(x)}M\psi(x)\nonumber \\
+{1\over 2}m^{2}_{0}(\rho^{\mu}_{i}\rho_{\mu i}+
\omega^{\mu}\omega_{\mu}+a^{\mu}_{i}a_{\mu i}+f^{\mu}f_{\mu})
\end{eqnarray}
where \(a_{\mu}=\tau_{i}a^{i}_{\mu}+f_{\mu}\), \(v_{\mu}=\tau_{i}
\rho^{i}_{\mu}+\omega_{\mu}\),
and \(u=exp\{i\gamma_{5}(\tau_{i}\pi_{i}+
\eta)\}\).
$u$ can be written as
\begin{equation}
u={1\over 2}(1+\gamma_{5})U+{1\over 2}(1+\gamma_{5})U^{\dag},
\end{equation}
where \(U=exp\{i(\tau_{i}\pi_{i}+\eta)\}\).
Since mesons are bound states solutions of $QCD$ they
are not independent degrees of freedom. Therefore,
there are no kinetic terms for meson fields. The kinetic terms
of meson fields are generated from quark loops. The scheme of nonlinear
$\sigma$ model is used to introduce the pseudoscalar meson fields.
Using the least action principle,
we obtain
\begin{eqnarray}
{\Pi_{i}\over\sigma}=i(\bar{\psi}
\tau_{i}\gamma_{5}\psi+ix\bar{\psi}\tau_{i}\psi)/(\bar{\psi}
\psi+ix\bar{\psi}\gamma_{5}\psi),\nonumber \\
x=(i\bar{\psi}\gamma_{5}\psi-{\Pi_{i}\over\sigma}
\bar{\psi}\tau_{i}\psi)/
(\bar{\psi}\psi+i{\Pi_{i}\over\sigma}
\bar{\psi}\tau_{i}\gamma_{5}
\psi),\nonumber \\
\rho^{i}_{\mu}=-{1\over m^{2}_{0}}\bar{\psi}\tau_{i}\gamma_{\mu}
\psi,\;\;\;
a^{i}_{\mu}=-{1\over m^{2}_{0}}\bar{\psi}\tau_{i}\gamma_{\mu}
\gamma_{5}\psi,\nonumber \\
\omega_{\mu}=-{1\over m^{2}_{0}}\bar{\psi}\gamma_{\mu}
\psi,\;\;\;
f{\mu}=-{1\over m^{2}_{0}}\bar{\psi}\gamma_{\mu}
\gamma_{5}\psi,
\end{eqnarray}
where \(u=e^{i\eta\gamma_{5}}(\sigma+i\gamma_{5}\tau\cdot\Pi)\)
, \(\sigma=\sqrt{1-\Pi^{2}}\), and \(x=tan\eta\).
The pseudoscalar fields have very complicated quark structures.

Substituting theses expressions into
the Lagrangian it becomes a Lagrangian of quarks.

Using the method of path integral to integrate out
the quark fields, the effective Lagrangian of mesons(indicated by M)
is obtained
\begin{equation}
{\cal L}^{M}_{E}=log det{\cal D},
\end{equation}
where
\begin{equation}
{\cal D}=\gamma\cdot\partial-i\gamma\cdot v-i\gamma\cdot a\gamma_{5}+mu.
\end{equation}
The Wess-Zumino-Witten Lagrangian with spin-1 fields is obtained from
the Lagrangian and
it is the leading term of ${\cal L}_{IM}$ in derivative expansion.

All the vertices of the meson physical processes of normal parity
can be found from ${\cal L}_{RE}$ and all the vertices of
abnormal parity can be derived from ${\cal L}_{IM}$.

{\large\bf Defining physical meson fields}\\
The physical meson fields can be defined in the following way
that makes
the corresponding kinetic terms in the standard form:
\begin{eqnarray}
\pi\rightarrow {2\over f_{\pi}}\pi,\;\;\;\eta\rightarrow
{2\over f_{\eta}}\eta, \nonumber \\
\rho\rightarrow {1\over g}\rho,\;\;\;
\omega\rightarrow {1\over g}\omega,
\end{eqnarray}
\begin{equation}
c={f^{2}_{\pi}\over 2gm^{2}_{\rho}}.
\end{equation}
\begin{equation}
a^{i}_{\mu}\rightarrow {1\over g}(1-{1\over 2\pi^{2}g^{2}})
^{-{1\over 2}}a^{i}_{\mu}-{c\over g}\partial_{\mu}\pi^{i}.
\end{equation}
\begin{equation}
f_{\mu}\rightarrow {1\over g}(1-{1\over 2\pi^{2}g^{2}})
^{-{1\over 2}}f_{\mu}-{c\over g}\partial_{\mu}\eta_{0}.
\end{equation}
In this theory the parameters are: \\
\hspace{3cm} {\large\bf $m_{u}$, $m_{d}$, $m_{s}$, g, m.}\\
g is an universal coupling constant. Input $f_{\pi}$, we determine
\(m=300MeV\) and choose \(g=0.35\).

{\large\bf Large $N_{c}$ expansion}\\
Due to the quark loops $N_{c}$ is included in this theory.
All tree diagrams are of order $O(N_{c})$, hence
they are leading contributions. A diagram with loops is at higher order
in large $N_{c}$ expansion. For instance, a diagram of one loop with
two internal lines is of order $O(1)$.
Therefore, the large $N_{c}$
expansion is the loop expansion in this theory.

{\large\bf Cut-off}\\
This theory is an effective theory and it is not renormalizable,
as mentioned above. Therefore,
a cut-off of momentum has to be introduced.
Using a cut-off instead dimensional regularization, we have
\begin{equation}
\frac{N_{c}}{(4\pi)^{2}}\{log(1+{\Lambda^{2}\over m^{2}})+
\frac{1}{1+\frac{\Lambda^{2}}{m^{2}}}-1\}={1\over 16}\frac{F^{2}}
{m^{2}}={3\over 8}g^{2}.
\end{equation}
Using the values of m and g, we obtain
\begin{equation}
\Lambda=1.6GeV.
\end{equation}
The masses of all mesons in this theory are below the cut-off.

{\large\bf Dynamical chiral symmetry breaking}\\
The quark condensate is defined as
\begin{equation}
<0|\bar{\psi}(x)\psi(x)|0>=-m^{3}{N_{C}\over 8\pi^{2}}
\{{\Lambda^{2}\over m^{2}}-log(1+{\Lambda^{2}\over m^{2}})\}.
\end{equation}
In nonperturbative $QCD$ a {\bf
dynamical quark mass}(constituent quark mass)is introduced
\begin{equation}
<0|\bar{\psi}(x)\psi(x)|0>=-m^{3}{3\over 4\pi\alpha_{s}}.
\end{equation}
Therefore, the parameter m of this theory is just the dynamical
quark mass in nonperturbative $QCD$ and
\begin{equation}
\alpha_{s}(1.6GeV)=\frac{4\pi}{{\Lambda^{2}\over m^{2}}-log(1+
{\Lambda^{2}\over m^{2}})}=0.50.
\end{equation}
There is dynamical chiral symmetry breaking in this theory.
On the other hand,
the PCAC is derived
\[\partial^{\mu}\bar{\psi}\tau_{i}\gamma_{\mu}\gamma_{5}\psi=-
m^{2}_{\pi}f_{\pi}\pi_{i}.\]

{\Large\bf Masses}\\
To the leading order in quark mass expansion,
the masses of the octet pseudoscalar mesons are derived(Gell-Mann,
Oakes, Renner formulas)
\begin{eqnarray}
m^{2}_{\pi}=-{2\over f^{2}_{\pi}}(m_{u}+m_{d})<0|\bar{\psi}\psi|0>,
\nonumber \\
m^{2}_{K^{+}}=-{2\over f^{2}_{\pi}}(m_{u}+m_{s})<0|\bar{\psi}\psi|0>,
\nonumber \\
m^{2}_{K^{0}}=-{2\over f^{2}_{\pi}}(m_{d}+m_{s})<0|\bar{\psi}\psi|0>,
\nonumber \\
m^{2}_{\eta}=-{2\over 3f^{2}_{\pi}}(m_{u}+m_{d}+4m_{s})
<0|\bar{\psi}\psi|0>.
\end{eqnarray}
\begin{equation}
m_{u}=4.64 MeV,\;\;\;m_{d}=8.16MeV,\;\;\;m_{s}=159MeV.
\end{equation}
The KSFR sum rule
\begin{equation}
g_{\rho}={1\over 2}f_{\rho\pi\pi}f^{2}_{\pi}
\end{equation}
is derived by using this theory and it
can be taken as the equation used to determine
$m_{\rho}$.
\begin{equation}
f_{\rho\pi\pi}={2\over g},
\end{equation}
\begin{equation}
g_{\rho}={1\over 2}gm^{2}_{\rho}.
\end{equation}
We obtain
\begin{equation}
m^{2}_{\rho}=6m^{2}.
\end{equation}
In the limit of \(m_{q}=0\),
the mass of $\rho$ meson originates
from dynamical chiral symmetry breaking.
\begin{equation}
m_{\rho}=0.751GeV.
\end{equation}
In the limit of \(m_{q}=0\),
the masses of the four low lying vector mesons originate from
dynamical chiral symmetry breaking.
\begin{equation}
f^{2}_{\pi}=3g^{2}m^{2}.
\end{equation}
The pion decay constant is the result of
dynamical chiral symmetry breaking too.
Therefore, in the limit of $m_{q}\rightarrow 0$,
the decay constants of the octet pseudoscalar mesons originate
from dynamical chiral symmetry breaking.

From the spontanous chiral symmetry breaking we obtain
\begin{equation}
(1-\frac{1}{2\pi^{2}g^{2}})m^{2}_{a}=6m^{2}+m^{2}_{\rho}.
\end{equation}
\begin{equation}
(1-\frac{1}{2\pi^{2}g^{2}})m^{2}_{f}=6m^{2}+m^{2}_{\omega}.
\end{equation}
Adding the strangeness to the Lagrangian, we obtain
\begin{eqnarray}
(1-\frac{1}{2\pi^{2}g^{2}})m^{2}_{K_{1}(1400)}
=6m^{2}+m^{2}_{K^{*}(892)}
\nonumber \\
(1-\frac{1}{2\pi^{2}g^{2}})m^{2}_{f_{1}(1510)}
=6m^{2}+m^{2}_{\phi}.
\end{eqnarray}

{\large\bf Summary of the results}\\
\begin{enumerate}
\item  Vector meson dominance(VMD)
\begin{eqnarray}
{e\over f_{\rho}}\{-{1\over 2}F^{\mu\nu}(\partial_{\mu}\rho^{0}_
{\nu}-\partial_{\nu}\rho^{0}_{\mu})+A^{\mu}j^{0}_{\mu}\},\nonumber \\
{e\over f_{\omega}}\{-{1\over 2}F^{\mu\nu}(\partial_{\mu}\omega_
{\nu}-\partial_{\nu}\omega_{\mu})+A^{\mu}j^{\omega}_{\mu}\},
\nonumber \\
{e\over f_{\phi}}\{-{1\over 2}F^{\mu\nu}(\partial_{\mu}\phi_
{\nu}-\partial_{\nu}\phi_{\mu})+A^{\mu}j^{\phi}_{\mu}\},
\end{eqnarray}
where $f_{\rho}={1\over 2}g$, $f_{\omega}={1\over 6}g$, and
$f_{\phi}=-{1\over 3\sqrt{2}}g$.
\item Weinberg's first sum rule is satisfied analytically.
\item The scattering lengths and slopes of $\pi\pi$ scattering obtained
by Weinberg are revealed from this theory. It is found that $\rho$ meson
exchange dominates $\pi\pi$ scattering.
\item In the chiral limits, two coefficients of chiral perturbation
theory are determined and in good agreement with CPT.
\item The amplitude of $\pi^{0}\rightarrow\gamma\gamma$ obtained in this
theory is the same as the triangle anomaly.
\item
The theory provides a
unified description of meson physics at low energies. In this
unified description universal coupling in all the physical processes
has been found and the inputs are the cut-off $\Lambda$, m(related to
quark condensate), and the quark masses. The theory is self-consistent
and phenomenologically successful.
\end{enumerate}

\newpage

\begin{table}[h]
\begin{center}
\caption { Summary of the results}
\begin{tabular}{|c|c|c|} \hline
    &  Experimental  &  Theoretical  \\ \hline
$f_{\pi}$   & 186MeV    & input         \\ \hline
g         &              & 0.35  input   \\ \hline
$m_{\omega}$& 781.94$\pm$0.12MeV    & 770MeV     \\ \hline
$m_{a}$     & 1230$\pm$40MeV    & 1389 MeV     \\ \hline
$m_{f_{1}}$     & $1282\pm5$MeV    & 1389 MeV     \\ \hline
$\pi$ form factor & consistent with $\rho$ pole & $\rho$ pole
\\ \hline
radius of $\pi$ & $0.663\pm$0.023fm & 0.63 fm        \\ \hline
$g_{\rho\gamma}$ & $0.116(1\pm 0.05)$ $GeV^{2}$
& 0.104 $GeV^{2}$ \\ \hline
$g_{\omega\gamma}$ & $0.0359(1\pm 0.03)$ $GeV^{2}$ & 0.0357 $GeV^{2}$
\\ \hline
$\Gamma(\rho\rightarrow\pi\pi)$ & $151.2\pm 1.2$ MeV & 135. MeV \\
\hline
$\Gamma(\omega\rightarrow\pi\pi)$ & $0.186(1\pm 0.15)$MeV & 0.136MeV
\\  \hline
$\Gamma(a_{1}\rightarrow\rho\pi)$ & $\sim$400 MeV & 325 MeV \\
\hline
$\Gamma(a_{1}\rightarrow\gamma\pi)$ &(640$\pm$246)keV & 252keV \\
\hline
${d\over s}(a_{1}\rightarrow\rho\pi)$ & $-0.11\pm 0.02$ &-0.097 \\
\hline
$\Gamma(\tau\rightarrow a_{1}\nu)$ &$(2.42\pm 0.76)10^{-13}$
GeV    &$1.56\times 10^{-13}$GeV   \\ \hline
$\Gamma(\tau\rightarrow \rho\nu)$ &$(0.495\pm 0.023)
10^{-12}$GeV   & 4.84$\times 10^{-13}$GeV \\  \hline
$\Gamma(\pi^{0}\rightarrow\gamma\gamma)$ & $7.74(1\pm 0.072) $eV
&7.64eV \\ \hline
a(form factor of $\pi^{0}\rightarrow\gamma\gamma$) & 0.032$\pm$
0.004 &0.03 \\ \hline
$\Gamma(\omega\rightarrow\pi\gamma)$ & 717(1$\pm$0.07)keV
& 724 keV \\ \hline
$\Gamma(\rho\rightarrow\pi\gamma)$ & 68.2(1$\pm$0.12)keV &
76.2keV \\  \hline
$\Gamma(\omega\rightarrow\pi\pi\pi)$ & 7.43(1$\pm$
0.02)MeV  &5 MeV \\ \hline
$\Gamma(f_{1}\rightarrow\rho\pi\pi)$ &
6.96(1$\pm$0.33)MeV&6.01MeV\\ \hline
$B(f_{1}\rightarrow\eta\pi\pi)$ &$(10^{+7}_{-6})\%$
&1.15$\times 10^{-3}$ \\ \hline
$\Gamma(f_{1}\rightarrow\gamma\pi\pi)$ &        &18.5keV  \\ \hline
$B(\rho\rightarrow\gamma\eta)$ &$(3.8\pm 0.7)\times 10^{-4}$
&3.04$\times 10^{-4}$\\ \hline
$B(\omega\rightarrow\gamma\eta)$ &$(8.3\pm 2.1)
\times 10^{-4}$ &$6.96\times 10^{-4}$ \\ \hline
\end{tabular}
\end{center}
\end{table}
\newpage
\begin{table}[h]
\begin{center}
\caption {Table II }
\begin{tabular}{|c|c|c|} \hline
    &  Experimental  &  Theoretical  \\ \hline
$m_{K_{1}}$     & 1402$\pm$7MeV    & 1510 MeV     \\ \hline
$m_{f_{1}(1510)}$     & 1512$\pm$4MeV    & 1640 MeV     \\ \hline
$g_{\phi\gamma}$ & $0.081(1\pm 0.05)$ $GeV^{2}$
& 0.086 $GeV^{2}$ \\ \hline
$<r^{2}>_{K^{\pm}}$ &$0.34\pm 0.05fm^{2}$ & $0.33 fm^{2}$     \\ \hline
$<r^{2}>_{K^{0}}$ &$0.054\pm 0.026fm^{2}$[11]
 & $0.0582 fm^{2}$   \\ \hline
$\lambda_{+}(K^{+}_{l3})$ & $0.0286\pm 0.0022$&0.0239
 \\ \hline
$\xi(K^{+}_{l3})$ &$-0.35\pm 0.15$ & -0.284\\  \hline
$\lambda_{+}(K^{0}_{l3})$ & $0.03\pm 0.0016$&0.0245
 \\ \hline
$\xi(K^{0}_{l3})$ &$-0.11\pm 0.09$ & -0.287\\  \hline
$\Gamma(K^{+}_{e3})$&$0.256(1\pm 0.015)10^{-17}$GeV&$0.233
\times 10^{-17}$GeV\\  \hline
$\Gamma(K^{0}_{e3})$&$0.493(1\pm 0.016)10^{-17}$GeV&$0.483
\times 10^{-17}$GeV\\  \hline
$B(\tau\rightarrow K^{*}(892)\nu)$ &$(1.45\pm 0.18)\%$
 &$1.46\%$   \\ \hline
$\Gamma(\tau\rightarrow K_{1}(1400)\nu)$ & &$0.373\%$ \\  \hline
$\Gamma(\phi\rightarrow K^{0}\bar{K}^{0})$ & $1.52(1\pm 0.03)$ MeV
& 1.11MeV \\ \hline
$\Gamma(\phi\rightarrow K^{+}K^{-})$ &$2.18(1\pm 0.03)$MeV & 1.7MeV \\
\hline
$\Gamma(K^{*}(892)\rightarrow K\pi)$ & $49.8\pm 0.8$MeV &39.4 MeV \\
\hline
$\Gamma(K^{+*}\rightarrow K^{+}\gamma)$&$50.3(1\pm 0.11)$keV&43.5keV\\
\hline
$\Gamma(K^{0*}\rightarrow K^{0}\gamma)$&$116.2(1\pm 0.10)$keV
&175.4keV\\ \hline
$B(K^{*}(892)\rightarrow K\pi\pi)$ & $0.53\times 10^{-4}
$ &$<7\times 10^{-4}$\\ \hline
$f_{1}(1510)\rightarrow K^{*}(892)\bar{K})$ & $35\pm 15 $MeV
&22.MeV \\ \hline
$\Gamma(K_{1}(1400)\rightarrow K^{*}(892)\pi)$ & $163.6(1\pm 0.14)$ MeV
 &126 MeV\\  \hline
$B(K_{1}(1400)\rightarrow K\rho)$ & $(3.0\pm 3.0)\%$
& $11.1\%$ \\ \hline
$B(K_{1}(1400)\rightarrow K\omega)$ & $(2.0\pm 2.0)\%$&
$2.4\%$ \\  \hline
$\Gamma(K_{1}\rightarrow K\gamma)$ &  &440keV\\ \hline
\end{tabular}
\end{center}
\end{table}
\begin{table}
\begin{center}
\begin{tabular}{|c|c|c|} \hline
$\Gamma(\eta'\rightarrow\eta\pi^{+}\pi^{-})$ &$87.8(\pm 0.12)$keV
&85.7keV\\ \hline
$\Gamma(\eta'\rightarrow\eta\pi^{0}\pi^{0})$ &$41.8(\pm 0.11)$keV
&48.6keV\\ \hline
$\Gamma(\eta\rightarrow\gamma\gamma)$ &$0.466(1\pm 0.11)$keV
&0.619keV\\ \hline
$\Gamma(\phi\rightarrow\eta\gamma)$ & $56.7(1\pm 0.06)$keV& 91.4keV
  \\ \hline
$\Gamma(\rho\rightarrow\eta\gamma)$ &$57.5(1\pm 0.19)$keV
&61.4keV\\ \hline
$\Gamma(\omega\rightarrow\eta\gamma)$ &$7.0(1\pm 0.26)$keV
 &7.84keV \\ \hline
$\Gamma(\eta'\rightarrow\gamma\gamma)$ &$4.26(1\pm 0.14)$keV
&4.88keV\\ \hline
$\Gamma(\eta'\rightarrow\rho\gamma)$ &$60.7(1\pm 0.12)$keV
&63.0keV\\ \hline
$\Gamma(\eta'\rightarrow\omega\gamma)$ &$6.07(1\pm 0.18)$keV
 &5.86keV \\ \hline
\end{tabular}
\end{center}
\end{table}

\newpage
{\large\bf Conclusions}\\
\begin{enumerate}
\item The masses of the octet pseudoscalar mesons are proportional to
current quark masses. In chiral limit, they are massless.
\item The masses of vector mesons originate from dynamical chiral
symmetry breaking.
\item The mass differences of vector and axial-vector mesons are caused
by spontaneous chiral symmetry breaking.
\item The cut-off $\Lambda$(1.6GeV) is at the border of nonperturbative
$QCD$ and perturbative $QCD$. Perturbative $QCD$ make corrections.
\item In the case of two flavors theoretical results agree with data
within about $10\%$ and if strange quark is involved, in the chiral
limit the worst case is $\Gamma(\phi\rightarrow K\bar{K})$ which is
less than data by $30\%$. The strange quark mass correction should
be taken into consideration.
\item The Lagrangian is not closed. Other mesons and glueballs, in
principle, could be included.
\item The parameter m and $\Lambda$ should be determined by full $QCD$.
\end{enumerate}

\end{document}